# A simple, sensitive and quantitative FACS-based test for SARS-CoV-2 serology in humans and animals

Agnès Maurel Ribes[1+], Pierre Bessière[2+], Jean Charles Guéry[3], Eloïse Joly Featherstone[4], Timothée Bruel[5], Remy Robinot[5], Olivier Schwartz[5], Romain Volmer[2], Florence Abravanel[3,6], Jacques Izopet[3,6] and Etienne Joly[7] *

1 : Laboratoire d'Hématologie, Centre Hospitalier Universitaire de Toulouse, 31000, Toulouse, France
2 : Ecole nationale vétérinaire de Toulouse, Université de Toulouse, ENVT, INRAE, IHAP, UMR 1225, Toulouse, France
3 : Institut Toulousain des Maladies Infectieuses et Inflammatoires (Infinity), Université de Toulouse, INSERM, CNRS, UPS, 31300 Toulouse, France
4 : York University Hospital, York YO31 8HE, UK
5: Institut Pasteur, Virus and Immunity Unit, CNRS-UMR3569, Paris, France
6: Université de Toulouse, INSERM, CNRS, UPS, 31300 Toulouse, France
7: Institute of Pharmacology and Structural Biology (IPBS), University of Toulouse, CNRS, Toulouse, 31000, France

+: AMR and PB made equivalent contributions to this work.
*: correspondence to EJ: atnjoly@mac.com

**Abstract**

Serological tests are important for understanding the physiopathology and following the evolution of the Covid-19 pandemic. Assays based on flow cytometry (FACS) of tissue culture cells expressing the spike (S) protein of SARS-CoV-2 have repeatedly proven to perform slightly better than the plate-based assays ELISA and CLIA (chemiluminescent immuno-assay), and markedly better than lateral flow immuno-assays (LFIA).
Here, we describe an optimized and very simple FACS assay based on staining a mix of two Jurkat cell lines, expressing either high levels of the S protein (Jurkat-S) or a fluorescent protein (Jurkat-R expressing m-Cherry, or Jurkat-G, expressing GFP, which serve as an internal negative control). We show that the Jurkat-S&R-flow test has a much broader dynamic range than a commercial ELISA test and performs at least as well in terms of sensitivity and specificity. Also, it is more sensitive and quantitative than the hemagglutination-based test HAT, which we described recently. The Jurkat-flow test requires only a few microliters of blood; thus, it can be used to quantify various Ig isotypes in capillary blood collected from a finger prick. It can be used also to evaluate serological responses in mice, hamsters, cats and dogs. FACS tests offer a very attractive solution for laboratories with access to tissue culture and flow cytometry who want to monitor serological responses in humans or in animals, and how these relate to susceptibility to infection, or re-infection, by the virus, and to protection against Covid-19.

**Introduction**

Over the past year, our world has been thrown into disarray by a pandemic caused by a new coronavirus, SARS-CoV-2. With a mortality rate around 1%, this new virus is not as pathogenic as previous coronaviruses such as SARS-CoV (9.6%) and MERS (35%), but it transmits faster from human-to-human (Fani et al. 2020), probably because of a large proportion (ca. 50%) of asymptomatic carriers (Wu et al. 2021; Long et al. 2020). Consequently, in less than two years since its discovery in Wuhan, China, SARS-CoV-2 has spread all over the world and caused more than 200 millions confirmed cases and over 4 millions confirmed deaths ( https://www.who.int/emergencies/diseases/novel-coronavirus-2019 ).

Biotechnology has proven a great asset in combating the pandemic, with the rapid development of diagnostic tests and, more recently, of vaccines. Diagnostic tests detect directly either the viral nucleic acid or viral proteins in nasopharyngeal swabs. Serological tests, by contrast, detect antibodies developed in response to infection by the virus or vaccination, or a combination of the two. Since the presence of antibodies in serum usually correlates with elimination of the virus and the patient's recovery, serological tests have not been very helpful in the clinic. They have, however, proven an essential tool to follow the spread of the pandemic by evaluating seroprevalence in populations, and they are now set to become essential to evaluate the immunity of individuals as well as populations (Koopmans and Haagmans 2020).

Whilst several thousands of serological studies have been published by now, based on tests performed in many millions of individuals, those myriad studies mostly document the seroprevalence in certain populations at a given time (Chen et al. 2021), but information regarding the actual protection afforded by immunity after infection by SARS-CoV-2, and how this correlates with the presence of antibodies against the SARS-2 virus is only just starting to come out (Lumley et al. 2021; Letizia et al. 2021; Abu-Raddad et al. 2021; Jeffery-Smith et al. 2021; R. A. Harvey et al. 2021; Garcia-Beltran et al. 2021). ).

While these recent publications show that the presence of antibodies does correlate with protection against the Covid-19 disease, and in particular against the more serious forms of the disease, one of the more burning questions that remains to be answered is how long this protection will last? Another crucial question concerns whether there will be differences in the duration of this protection depending on which vaccine was used, and whether an individual had been infected by the SARS-CoV-2 virus before, or after, being vaccinated.

Obtaining answers to this type of questions should be greatly facilitated by access to simple, cheap and quantitative serological tests, which would work both in humans and in animal models. To date, however, although a multitude of commercial serological tests have been developed to detect the presence of antibodies in the serum of patients (Farnsworth and Anderson 2020), those are mostly ill-suited for use in research laboratories, not only because of their price, but also because they are not or only poorly quantitative.

The most commonly used methods for Covid-19 serodiagnostic are either ELISA (Enzyme-Linked ImmunoSorbent Assays) or CLIA (ChemiLuminescent ImmunoAssays). Whilst those methods show very good sensitivity and specificity, they also have several significant drawbacks:

i) The commercial versions are based on using volumes of serum or plasma which exceed the amounts which can be readily obtained by finger prick, and therefore require venipuncture, and hence trained personnel to collect the samples, and elaborate logistics to handle those samples.
ii) They are relatively expensive (ca. 500 € per plate of 90 tests for commercial ELISA or CLIA ) and not easily modular (i.e. a whole plate will often have to be used even if only a few tests are to be performed). Whilst in-house ELISAs are a possible alternative, they are difficult to standardize and require high amounts of recombinant antigen (Amanat et al. 2020).
iii) Whilst ELISA tests are quantitative, they tend to saturate rapidly, and thus show a relatively limited dynamic range.
iv) Most commercial versions are designed to detect human antibodies, and thus cannot be used to follow serological responses in animal models.

Early in the pandemic, because they could be used in a point of care setting on capillary blood obtained by finger pricks, lateral-flow immune assays (LFIA) attracted considerable attention as an alternative to the ELISA or CLIA plate-based methods. Dozens of versions were developed by various commercial companies, and despite their relatively high price, such LFIAs were used in scores of studies to evaluate the prevalence of sero-conversion in various populations. Over time, however, the general performances of LFIAs have proven to be too low, both for sensitivity and reliability, to be of real use in clinical settings, and even for epidemiological studies (Adams et al. 2020; Mohit et al. 2021; Dortet et al. 2021; Moshe et al. 2021).

As an alternative to those various serological tests, we set out to develop a serological test based on hemagglutination, with the aim of obtaining a method that would be both sensitive, cheap, and could be used both in the laboratory, in field settings or as a point of care test, without the requirement for any elaborate equipment. The HAT (HemAgglutination Test) method is based on a single reagent, IH4-RBD, which binds to human red blood cells (RBC) via the IH4 nanobody specific for human Glycophorin A and coats them with the RBD domain of the SARS-CoV-2 virus (Townsend et al. 2021). HAT has a sensitivity of 90% and specificity of 99%, and is now used by several laboratories worldwide for epidemiological and clinical studies (Kamaladasa et al. 2021; Jayathilaka et al. 2021; Jeewandara et al. 2021; Ertesvåg et al. 2021). To be able to perform HAT on whole blood with the sensitivity and simplicity which we set out the reach, and to attempt to make it quantitative, various modifications and improvements had to be tested, and in order to do this, we felt that we needed a simple quantitative test that would allow us to evaluate the amount of antibodies present in the whole-blood samples we were using more simply and cheaply than by using ELISA or CLIA.

In this regard, the S-flow test (Grzelak et al. 2020), which uses flow cytometry (FACS) performed on human cells expressing the S protein, appeared as a very promising approach since it is very simple to run and its performances compared favorably with three other serological tests (two ELISA directed towards the S or N proteins, and a luciferase immuno-precipitation system (LIPS) combining both N and S detection).

The S-flow method initially described used HEK cells (Grzelak et al. 2020), which are adherent cells. We felt that it would be better to use cells growing in suspension, not only because it makes it a lot easier to grow large numbers of cells, but also because those can be used directly, without having to be detached from the plastic, which we feared could possibly alter the cells' characteristics, and introduce a possible source of variability between assays.

In this regard, Horndler and colleagues have recently described an assay inspired by the S-flow assay, but based on Jurkat cells expressing both the full-length native S-protein of SARS-CoV-2 and a truncated form of the human EGFR protein, which is used as an internal control for the expression level of the S protein (Horndler et al. 2021).

One of the caveats of using human cells to express the S protein, however, is that those cells will also express other antigens, and MHC molecules in particular, which can be the targets of allo-reactive antibodies present in certain individuals, and not others. The background level of staining on Jurkat cells themselves will thus vary from sample to sample. To circumvent this difficulty, Pinero et al. have used the Jurkat-S+EGFR cells mixed with untransfected Jurkat cells as negative controls (Piñero et al. 2021).

Having followed the same reasoning as Horndler *et al.*, we had chosen to make use of Jurkat cells expressing the native form of the spike protein of the SARS-CoV-2 virus, but we elected to use Jurkat cells expressing the mCherry red fluorescent protein as negative controls. This is not only cheaper because it does not require labelling the cells with an additional commercial antibody, but also has the advantage of using two cell lines that are maintained in the same selective tissue culture medium.

In all three papers (Horndler et al. 2021; Piñero et al. 2021; Grzelak et al. 2020), and several others (Egia-Mendikute et al. 2021; Hambach et al. 2021; Lapuente et al. 2021; Goh et al. 2021), the sensitivity of approaches based on flow cytometry was reported to be superior to those of ELISA or CLIA, probably because such tests are based on detecting the spike protein expressed in its native conformation.

Since the Jurkat-flow test calls for the use of both a flow cytometer and cells obtained by tissue culture, it is clearly not destined to be used broadly in a diagnostic context, but its simplicity, modularity, and performances both in terms of sensitivity and quantification capacities should prove very useful for research labs working on characterizing antibody responses directed against SARS-2, both in humans and animal models.

## Results and Discussion

### *Jurkat-S&R-flow: basic principles*

Jurkat cells expressing high levels of the SARS-CoV-2 spike protein, which we subsequently refer to as Jurkat-S, or J-S, were obtained by means of transduction with a lentiviral vector, followed by three rounds of cell sorting. A second population of Jurkat cells, in which all cells express the mCherry fluorescent protein, and which we subsequently refer to as Jurkat-R, or J-R, was also obtained by lentiviral transduction (see M&M: Materials and Methods section).

For the Jurkat-S&R-flow test, we simply prepare a mix of equivalent numbers of the two cells lines, J-S and J-R, and use either sera or plasma at a final dilution of 1/100 to label $2.10^5$ cells of this mix (see M&M for details). After this primary step of labelling, the cells are washed before being incubated with a fluorescent secondary antibody, and a final wash is performed before analysis by flow cytometry.

The advantage of using such an approach is that it guarantees that the test cells (Jurkat-S) and the control cells (Jurkat-R) are labelled in exactly the same conditions. Comparing the levels of staining between test and control cells can then be carried out without any risk of a difference between the two being due to a difference in the course of the labelling procedure (e.g. certain samples receiving less of the primary or secondary antibodies). Accessorily, another significant advantage of such an approach is that it reduces the number of FACS samples to be processed by a factor of two.

The mCherry signal allows simple separation by gating during analysis of the control cells from the test cells (Figure 1). In samples labelled just with the secondary antibody (neg. cont.) or with pre-pandemic plasma which does not contain antibodies against the spike protein (neg plasma), similar green signals are found on J-S et J-R populations. When there are antibodies against the spike protein present in the plasma or serum used to label the cells, this will result in a marked difference in the green signals detected on the J-S cells compared to the J-R cells. The difference in the fluorescence intensity between the two populations will provide a quantitative evaluation of the amounts of antibodies in the serum or plasma used to label the cells (first column). The numbers shown in red correspond to the relative specific staining (RSS = signal J-S – signal J-R / signal neg. cont.).

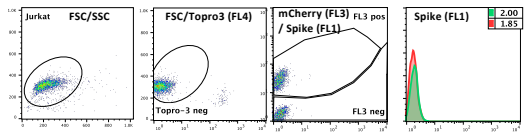
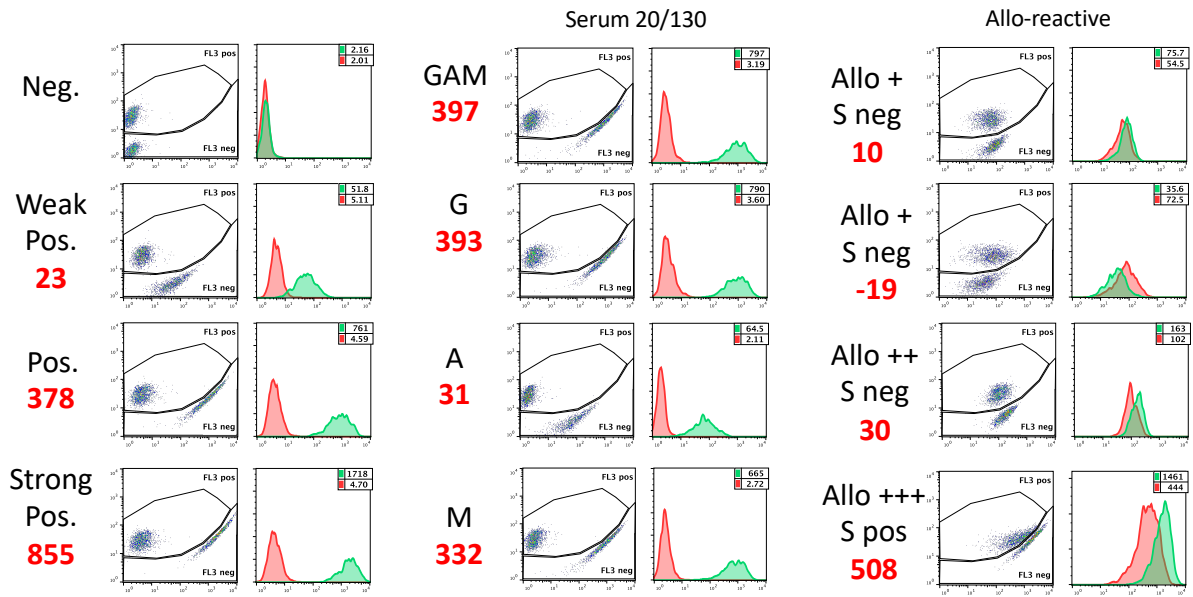

**Figure 1**: Basic principle and examples of the Jurkat-S&R-flow test

A 50/50 mix of Jurkat cells expressing either the S protein of the SARS-CoV-2 virus or the mCherry protein is first incubated with a 1/100 dilution of the plasma or serum sample to be tested, followed by incubation with a green fluorescent secondary antibody. The cells are then analyzed by flow cytometry.

The panels on the first line show the gating strategy, using as an example the sample of a negative control, stained only with an anti-human pan specific secondary antibody conjugated to alexa-488 : Live cells are first selected on a combination of two gates drawn on three parameters: size (FSC), granularity (SSC), and far red fluorescence (FL4) to eliminate the dead cells labelled by the Topro3 live stain. Jurkat-S are then distinguished from Jurkat-R cells using the red fluorescence of the latter, by means of a gate drawn on a FL1/FL3 dotplot. The gates had to be drawn that way to accommodate the fact that, for the brightest cells, the green fluorescent signal of alexa-488 can 'bleed' into the FL3 channel, and the Cellquest software does not allow for FL3/FL1 compensation. A histogram overlay is then drawn with the cells falling in each of these two gates (red: Jurkat-R, green: Jurkat-S). The numbers in the upper right corners of those histograms correspond to the GMFIs of the two histograms.

The red numbers shown to the left of the plots correspond to the relative specific staining (RSS; i.e. the difference of signal between J-S and J-R divided by the GMFI of the negative control, i.e. 2.00)

The first column shows examples of staining with 4 different plasmas which were either negative, or weakly, positively and strongly reactive with the spike protein.

The second column shows an example of the Jurkat-S&R-flow test capacity to perform a rough evaluation of the relative proportion of the various Ig isotypes in a sample, in this case the 20/130 reference serum obtained from the NIBSC.

The third column presents examples of some of the rare samples which showed various levels of allo-reactivity against the Jurkat cells themselves. Although the sample on the third line had an RSS of 30 (thus well above the threshold of 20 set for positivity), this sample was probably negative since it showed no reactivity in the other two serological tests. The sample on the fourth line shows that some sera can be both allo-reactive against the Jurkat cells, and strongly reactive against the spike protein.

If isotype-specific secondary reagents are used, an evaluation of the respective amounts of various classes of antibodies can also be obtained (second column). It should, however, be noted that, because different secondary reagents do not necessarily recognize the various Ig isoforms with the same efficiency, this provides only a very rough analysis of the relative amounts of Ig-G, -A and -M. Within a set of samples, however, this can provide very simple means to compare samples with one another (Table S1).

As alluded to in the introduction, one of the possible caveats of using a human cell line to express the S protein is that some blood samples will contain allo-reactive antibodies directed against that cell line, possibly as a consequence of a pregnancy, or past history of receiving a blood transfusion or organ transplant (Hickey et al. 2016; Karahan et al. 2020). The third column of Figure 1 shows examples of such samples containing marked levels of alloreactive antibodies, i.e. samples for which the J-R cells show significant levels of staining compared to the same cells labelled with just the secondary antibody. Based on our results collected on more than 350 clinical samples, we evaluate that ca. 30 % of samples will contain allo-reactive Abs that will result in levels of staining of Jurkat cells that are more than five-fold that of the signal obtained for the negative control (and 3-6 % more than ten-fold). Of note, we did not notice an increased frequency of allo-reactivity in samples from women compared to men, which suggests that allo-reactivity after pregnancy is not a major cause in the origin of those allo-reactions.

A difficult question with all serological tests is that of where to set the threshold beyond which the specific signals detected can confidently be considered as positive, which will be directly linked to the balance between sensitivity and specificity of the assay. Based on the analyses of various cohorts of positive and negative samples (some of which will be presented further down in this manuscript), for the Jurkat-S&R-flow, we have settled for a threshold of RSS =20, i.e. twenty-fold the value of the negative control stained just with secondary-antibodies. The reason for using the geometric mean of fluorescence intensity (GMFI) of the negative control as an internal reference is that, in flow cytometry, the numerical values of fluorescence intensities will be totally dependent upon the cytometer settings, and the voltages applied to the PMT in particular. But we find that this can be somewhat compensated by such an approach. For example, in the conditions used in our experiments, the GMFI of the negative control had a value of ca. 2 when analyzed on a FACScalibur flow cytometer. With a threshold set at 20, samples were thus considered as positive if the difference in the GMFI of the J-S and J-R populations was above 40 (numbers shown in red in Figure 1 are J-S – J-R / 2). When the very same samples were analyzed on a Fortessa flow cytometer (Figure S1), the value of the GMFI for the negative control was 24-fold higher, but the RSS values obtained closely resembled those obtained with the same samples on the FACScalibur (red numbers in Figure S1 and Figure 1: 20/23; 366/378; 916/855; 490/508).

For the samples showing high levels of allo-reactive staining, however, it is worth underlining that this threshold of 20 had to be considered with some caution. Indeed, in those allo-reactive samples, such as the examples shown in the right column of Figure 1, we found that the difference between the green signals recorded for the J-S and J-R populations can fluctuate between experiments, presumably because those signals correspond to the recognition by allo-reactive antibodies of cell surface markers that are not always expressed at the same levels in all Jurkat cells. Such alloreactive signals can be higher in J-R for certain serum or plasma samples, or in J-S for other samples (first vs second and third line), without, in this latter case, necessarily corresponding to bona-fide reactivity against the spike protein of the SARS-CoV2 virus. The sample shown on the fourth line represents the most extreme case we have come across, from a Covid-19 patient with extremely high allo-reactivity, coupled to bona fide reactivity against the spike protein. All in all, it is simply worth underlining that, for the small percentage of samples showing significant alloreactivity against Jurkat cells, and staining of Jurkat-S slightly higher than that of Jurkat-R, their reactivity should be checked with a different serological test before considering them as truly positive.

Inspection of the dot plot for this very alloreactive sample provides the explanation for the slightly odd shapes of the FL1/FL3 gates we used to discriminate Jurkat-S from Jurkat-R cells. This was necessary because, on the FACScalibur, the fluorescent signal of mCherry is best captured by the FL3 channel, but the version of the Cellquest program used for acquisition on this cytometer does not allow for FL3/FL1 compensation. Consequently, samples with extremely high FL1 signals showed some 'bleeding' into the FL3 channel. We found that this could not be satisfactorily treated by post-acquisition compensation with the Flowjo analysis software either, and thus resorted to drawing such gates to separate J-S from J-R populations.

We had elected to use a green/red combination for test and control cells for two reasons: 1) secondary antibodies labeled with green fluorescent dyes such as fluorescein or Alexa 488 are the most commonly available, and also usually the cheapest. 2) All flow-cytometers, even the most basic ones, are equipped with a 488 nm laser which allows to perform green/red analyses. Another important consideration is that one of the goals of this study was to set up a test which could be used by as many research teams as possible, including those based in institutes from less affluent countries, which are less likely to have access to recent, state-of-the-art multi-laser flow cytometers. As will be seen later, with secondary antibodies conjugated to Alexa-647 which can be excited by the second 633 nm laser of the FACScalibur, it is possible to use Jurkat-GFP cells as an alternative to J-R cells.

When the same samples as shown on Figure 1 were analyzed using a Fortessa flow cytometer, the picture was quite different (Figure S1). This more recent flow cytometer is endowed with several lasers, including a 561 nm Yellow-Green laser, which is much better suited for the excitation of the mCherry fluorescent protein, thus yielding much higher signals which, since they are acquired on a different laser line from the green signals, require absolutely no compensation.

In many flow cytometry facilities, users are required to fix any samples that have been in contact with materials of human origin. Whilst this was not the case for us, we still tried analyzing the same samples after those were fixed with 1% formaldehyde and kept for 4 days at 4°C before re-analysis. As can be seen on the right part of Figure S1, formaldehyde-fixation resulted in a 7-fold drop in the intensity of the mCherry signals, which made it impossible to separate J-R from J-S populations on the FACScalibur (first line). On the other hand, analysis on the Fortessa was still comfortably possible. Of note, whilst formaldehyde did not noticeably alter the fluorescence signals of the alexa-488 dye, it did result in a twofold increase of the green auto-fluorescence of the negative controls, hence resulting in a twofold reduction of the RSS compared to those obtained on unfixed cells.

*Performance of the JurkatS&R-flow test on clinical samples*

Next, we compared the performance of the Jurkat-S&R-flow test with those of two other serological tests: the Wantai commercial RBD ELISA test, and the hemagglutination-based test, HAT (Townsend et al. 2021).

For this, we made use of two cohorts of clinical samples.
First, a preexisting cohort of 121 sera available in the virology laboratory of the Toulouse hospital which had all previously been tested with the Wantai RBD ELISA test (Abravanel et al. 2020), comprising 40 negative serum samples collected before December 2020, and 81 sera from PCR-positive subjects. For those samples, RBCs from O- donors were used to perform the HAT tests (see M&M).
The second cohort consisted of 267 whole blood samples collected on EDTA, obtained from the hematology department of the Toulouse hospital as left-over clinical material, for which we have consequently very little clinical information. For those samples, the HAT test was performed on whole blood, i.e. using the subjects' own RBCs for hemagglutination. The blood samples were subsequently centrifugated, and the plasmas collected to perform the Jurkat-S&R-flow and the RBD-ELISA tests.

The result of the analysis of these two cohorts with the three serological tests are presented in Figure 2, with colors used to represent the HAT results.

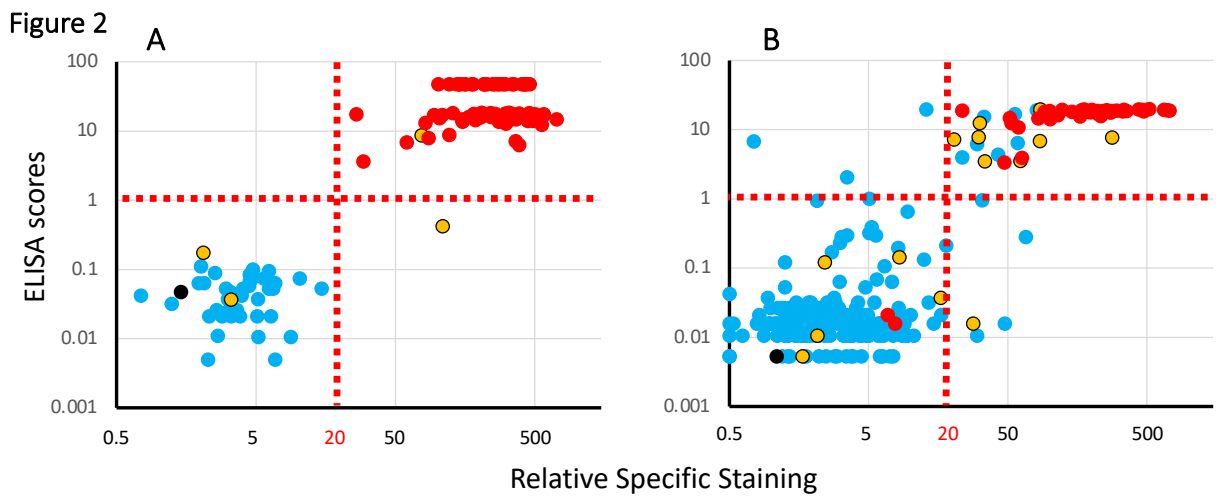

**Figure 2**: Comparison of the results of the 3 serological tests on two cohorts of clinical samples.
Panel A: results obtained with a cohort of 121 serum samples (81 PCR pos, 40 neg) collected by the virology department of the Purpan Hospital (Toulouse, France).
Panel B: results obtained with a cohort of 267 whole blood samples coming from the hematology department of the Rangueil Hospital (Toulouse, France).
Each plot presents ELISA scores (Y axis) against the results of the Jurkat-S&R-flow tests, expressed as RSS. For improved clarity, both axes are presented on log scales (for this, the negative RSS values of 4 samples of the cohort on the right had to be manually converted to a value of 0.5).
Colors are used to represent the HAT results: red: positive, blue: negative, yellow: partially positive hemagglutination, black: false positive, i.e. samples for which hemagglutination also occurred with the IH4 nanobody not coupled to the RBD domain.

The results of the first cohort (Figure 2A) show a clear-cut dichotomy in the distribution of the points, with two well separated clouds: one of blue points falling in the lower left quadrant, and one of red points in the upper right one. In other words, for this cohort of sera collected either from hospitalized Covid-19 patients or dating from before the pandemic for the negative controls, the three tests are in almost perfect agreement for the discrimination between positives and negatives, with just one yellow point in the lower right quadrant, i.e. positive for the Jurkat-S&R-flow test, showing only partial hemagglutination, but the signal for the RBD-ELISA test falling slightly below the threshold of 1 set by the manufacturer, suggesting that the anti-viral serological response of this subject was probably focused on regions of the spike protein outside of the RBD domain.

Of note, partial hemagglutination was also recorded for two other samples which were negative for the two other tests. Those two samples were from a group of 38 patients hospitalized for Covid-19, but for whom the blood had been collected less than 14 days after the PCR diagnostic. This observation is in line with our previous observation that the HAT test may be particularly performant for the detection of early serological responses, probably because of a higher hemagglutinating capacity of IgMs (Townsend et al. 2021). Incidentally, in this same group of 'early' PCR-positive Covid patients, there were also 4 other samples which were negative for all three tests (see tables of data provided as supplementary material).

The results of the second cohort, which comprised a few Covid patients, but also a large proportion of blood samples randomly picked among those from patients hospitalized for conditions unrelated to Covid-19, yielded a much less clear picture than the first one. As can be seen on Figure 2B, many of the dots for this cohort occupy a more intermediate position between the two clouds of clearly positive and clearly negative samples. In the upper right quadrant, one notices a relatively high proportion of blue and yellow spots, i.e. of HAT negative or partial samples in the set which were positive with both the RBD-ELISA and Jurkat-S&R-flow tests, albeit in the lower left part of the quadrant, i.e. rather weakly.

On the other hand, there are also a handful of samples in the lower left quadrant for which partial or full hemagglutination was detected, which may correspond to early serological responses. There are also a few blue dots in the upper left and lower right quadrant, thus corresponding to samples being positive only either with the RBD ELISA or the Jurkat-S&R-flow test. According to previous work, the sera which react weakly on the full length spike protein expressed at the surface of Jurkat cells are most likely due to cross-reactivities with the S2 domain of other coronaviruses (Ng et al. 2020; Khan et al. 2020).

One additional conclusion that can be drawn from the comparison of the results of the RBD-ELISA with those of the Jurkat-S&R-flow test is that, whilst the two methods show similar sensitivities, the ELISA signals tend to saturate very rapidly, and are thus much less dynamic that those obtained by flow cytometry.

*Examples of possible uses of the Jurkat-flow test*

A further advantage of the Jurkat-S&R-flow test compared to the commercial RBD-ELISA we used for this study is that it only requires 1 µL of plasma or serum, compared to 100 µL for the ELISA, which makes it possible to perform on small volumes of capillary blood collected by finger prick.

As a proof of concept that this was doable and useful, we used capillary blood which one of the authors collected by pricking his finger at various intervals to document the time course of his serological response after vaccination, and the results are shown in panel A of Figure 3. One somewhat surprising finding was with the low levels of IgM recorded, which may in part be explained by the fact that, as a rule, anti-IgM secondary antibodies tend not to work as well as those against the other isotypes. One should note, however, that, with the same anti-IgM secondary reagent, signals of the same order of magnitude were found for IgG and IgM with the 20/130 reference serum (Figure 1), as well as in several other samples (Table S1). The observation that the signals obtained with an anti-IgG reagent are similar, and even often a bit higher to those obtained with the pan-reactive anti-IgGAM is something which we tend to find in most samples (Table S1), and which had already been underlined by Grzelak et al. in the context of the S-flow test (Grzelak et al. 2020).

The capacity of the SARS-CoV-2 virus to infect primates, hamsters or ferrets has provided very useful animal models to understand the physiopathology of Covid-19, whilst its capacity to infect domestic pets such as cats and dogs does raise concerns about the transmissions of the virus back to humans, as well as about the problem of animal reservoirs from which it will be very difficult to eradicate the virus. Whilst scores of commercial and lab-made methods have been described to follow human serological responses to the virus, there is a remarkable paucity of tests available today applicable to animals. Since all it would take to adapt the Jurkat-S&R-flow test to animals would be to use different species-specific secondary antibodies, we explored whether this would work for mice, cats, dogs and hamsters.

Figure 3

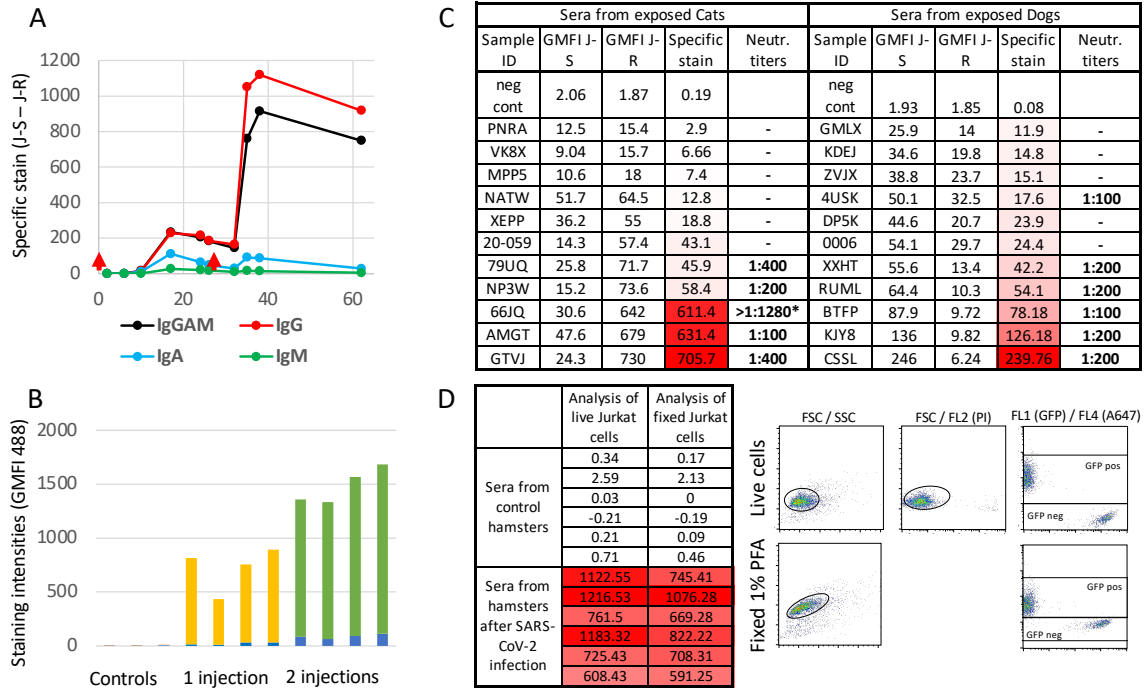

**Figure 3**: Various examples of the possible uses of the Jurkat-flow test

**Panel A**: Monitoring serological responses in humans
   The Jurkat-S&R-flow test was used to follow the serological response of an individual during the course of his vaccination (red arrows: first injection on day 0 and boost on day 27), with isotyping using specific secondary reagents performed as described in the M&M section. Y axis: Specific staining ( GMFI J-S – GMFI J-R)

**Panel B**: The Jurkat-S&R-flow test can be used to follow Ig responses in mice.
   Sera from three groups of mice having been immunized either once, twice or not by intra-peritoneal injection of inactivated SARS-CoV-2 virus were analyzed with the Jurkat-S&R-flow test, using an anti-mouse Ig secondary antibody. Y axis: GMFI on Jurkat-R (blue columns), on Jurkat-S (top of green columns), The specific signals correspond to the yellow or green portions (GMFI J-S – GMFI J-R).

**Panel C**: In cats and dogs, the results of the Jurkat-S&R-flow test correlate with those of sero-neutralization titers.
   Panels of sera from cats and dogs whose owners had recovered from symptomatic Covid-19 were used to perform both sero-neutralization assays and the Jurkat-S&R-flow test, using the corresponding secondary antibodies. Specific stain = GMFI J-S – GMFI J-R.  *: the higher SN titers shown for the cat serum 66JQ was obtained on a different day from all the others.

**Panel D**: Monitoring of antiviral serological responses in Hamsters with the Jurkat-S&G-flow test
   For this experiment, we used sera from 6 hamsters which had been experimentally infected with the UCN1 strain of SARS-CoV-2 15 days earlier, as well as 6 uninfected controls. Since we had an anti-hamster Ig secondary antibody conjugated to alexa 647, we used an S&G version of the test, with Jurkat cells expressing GFP as negative control, and propidium iodide to gate out dead cells (see M&M). As can be seen on the right side of the panel, in this configuration, no compensation was required, and separating the Jurkat-S test cells from the Jurkat-G control cells could be performed with simple square gates. And although the GFP signal was reduced after fixation of the cells in PFA for 24 hours analysis of cells was still possible with the FACScalibur, with almost no loss of the alexa 647 signal.

First, we used sera of mice which had been immunized once or twice with inactivated SARS-CoV-2 virus injected intra-peritoneally. As can be seen on panel B of Figure 3, whilst the sera of control mice did not react significantly with Jurkat cells, those of immunized mice showed strong specific reactions against the spike protein, which were even higher in those having received two injections. Of note, the reactions against the Jurkat-R cells also went up in correlation with the injections, albeit to a much lesser degree than against the S protein. This is presumably due to the fact that the preparations of inactivated virus used for the immunizations probably contained some bovine proteins from the serum used in the tissue culture medium that would have the capacity to bind to the surface of the Jurkat cells, such as, for example, beta-2-microglobulin binding to MHC class I molecules. The Jurkat-S&R-flow method therefore seems to work very satisfactorily in mice, and since it requires only a few µL of blood, it is well suited to follow serological responses over time.

We then turned our attention to cats and dogs, which have both been shown to be susceptible to infections by the SARS-CoV-2 virus (Sit et al. 2020; Dróżdż et al. 2021; Bessière, Fusade-Boyer, et al. 2021; Chen 2020), but for which one of the only reliable means to test for the presence of a serological response is to perform sero-neutralization experiments, which are both cumbersome and require access to BSL-3 facilities. For our exploratory experiments, we simply used sets of sera collected from 11 cats and 11 dogs whose owners had had symptomatic Covid-19 infections, and on those we performed both sero-neutralization experiments, and the Jurkat-S&R-flow test, using the appropriate secondary antibodies. As can be seen on panel C of Figure 3, we found a very good correlation between the levels of specific staining of Jurkat-S cells and the neutralization titers of the same sera. Whilst those are very preliminary observations which will need to be strengthened by many more samples, and in particular with samples collected before the Covid-19 pandemic as negative controls, those results show that the Jurkat-S&R-flow test can clearly be used to evaluate the levels of antibodies against the S protein of the SARS-CoV-2 virus in cats and dogs.

Finally, we turned our attention to hamsters, which are susceptible to infection by SARS-CoV-2, and have been a very useful animal model to study the physiopathology of the infection. For this, we made use of a panel of sera from hamsters which had been used in a previous study (Bessière, Wasniewski, et al. 2021). Because we had some anti-hamster Ig conjugated to the Alexa -647 fluorochrome at our disposal, we took this opportunity to test the possibility of adapting the Jurkat-S&R-flow test to a Jurkat-S&G-flow test, i.e. using Jurkat cells expressing GFP rather than mCherry as the internal negative control. As can be seen on panel D of figure 3, the sera from the hamsters that had been experimentally inoculated with the SARS-CoV-2 virus two weeks earlier all harbored very high levels of antibodies against the virus, whilst no signal was detected with the sera from control animals. Of note, with the S&G version of the test, i.e. using GFP-expressing Jurkat cells as negative control and a secondary antibody conjugated to alexa-647, since the two signals were collected on the separate green and red laser lines of the FACScalibur, the situation was similar to when we analyzed the Jurkat-S&R-flow test on a Fortessa flow cytometer: gating to separate the test from the control cells was achieved with simple rectangular gates, and although fixation in PFA resulted in a twofold reduction of the GFP fluorescent signal, and the auto-fluorescence of the Jurkat-S cells increased about threefold, the efficient triggering of GFP by the 488 nm laser meant that the test and control populations could still be separated and analyzed after fixation, albeit staining with propidium iodide could no longer be used to gate out dead cells.

## Concluding remarks and perspectives

Given its versatility, flexibility and affordability, we believe that the Jurkat-S&R-flow or Jurkat-S&G-flow assays could prove useful for many research scientists wanting to measure serological responses directed towards the spike protein of the SARS-CoV-2 virus, either in humans, or in animals, but who would not have either the financial means to purchase commercial plate-based assays such as ELISA or CLIA, or access to the sizeable amounts of recombinant proteins required to set those up in-house. The Jurkat-flow tests could be performed by any laboratory with access to tissue culture and to a flow cytometer, at a cost of roughly 10 cents per sample (see M&M for calculation). It is thus much cheaper than commercial ELISAs, which cost around 500 € per plate of 90 samples. And the Jurkat-flow test is also completely modular, i.e. each test only comprises the number of samples required, contrary to plate-based assays for which it is rather difficult not to use up a whole plate every time.

Several reports have already highlighted that using flow-cytometry for the detection of antibodies binding to the S protein expressed at the surface of cells tends to perform better than plate-based assays which use immobilized recombinant proteins (Egia-Mendikute et al. 2021; Hambach et al. 2021; Lapuente et al. 2021; Goh et al. 2021; Horndler et al. 2021; Piñero et al. 2021; Grzelak et al. 2020; Ng et al. 2020). This is most likely related to the fact that many antibodies will be binding to conformational epitopes, which will only be found on the naturally expressed and properly folded spike protein. In this regard, the capacity of the S protein to undergo structural fluctuations, particularly at the level of the RBD domain which can be in either an open/up or closed/down conformation, has been shown to influence the binding-capacity of various antibodies (W. T. Harvey et al. 2021; Barnes et al. 2020; Robbiani et al. 2020; Piccoli et al. 2020; Zhou et al. 2020; Huang et al. 2021; Dejnirattisai et al. 2021). This is supported by the results shown in Figure S2, i.e. that, for most samples, we found that incubation at room temperature resulted in a sizeable increase in the amounts of antibodies binding to the spike-expressing cells, suggesting that, at the surface of live cells, the capacity of the spike protein to fluctuate between various conformations will expose different epitopes, and allow the binding of more antibodies.

One striking aspect of the results shown on Figure 2 is in the difference of performance of the tests between the two cohorts of samples tested. On the one hand, nearly perfect scores were obtained for all three tests on a cohort of sera comprised either of control samples collected before 2019, or of positive sera from PCR-positive Covid-19 patients. On the other hand, the situation was much less clear-cut for the cohort comprising blood samples picked more or less randomly and blindly among those available as left-overs from the hematology department and was, therefore, more akin to a 'real' population. For this second cohort, an additional confounding factor may have been that, since all the samples were from hospitalized patients, some sera may have been poly-reactive due to inflammatory pathologies unrelated to Covid-19.

All in all, this difference between the two cohorts is reminiscent of the common observation that the performances of clinical tests are often much lower on real populations than those obtained by the manufacturers on very carefully controlled and standardized populations. Our results indeed bring support to the view that the performance of any given serological test will be entirely dependent on the set of samples used to measure it: whilst it is relatively easy to reach an almost perfect score on a cohort comprised only of highly positive and completely negative samples such as the one used for Figure 2A, the situation becomes much less clear when using a set of samples more closely resembling the general population, in which the positive or negative nature of many samples will remain uncertain, and from which it would thus seem futile to try to make precise calculations of sensitivity and specificity. This being said, performing several tests in parallel on such cohorts is very useful to compare the performance of those tests with one another.

Several recent reports have underlined the correlation between the serological levels of neutralizing antibodies, which are mostly directed against the RBD, and the degree of protection against becoming infected, or re-infected, by the SARS-Cov-2 virus (Feng et al. 2021; Khoury et al. 2021; Garcia-Beltran et al. 2021). Low levels of antibodies are often found in convalescent subjects following asymptomatic infections, and Kalamadasa and colleagues have shown that, whilst HAT sensitivity in such individuals can be as low as 50 %, HAT results are strongly correlated with the seroneutralisation activity in those samples (Kamaladasa et al. 2021). Over the coming months and years, antibodies levels will progressively decrease in both vaccinated and convalescent people, and one of the crucial questions will be that of when to start planning to administer vaccine boosts. Levels of neutralizing antibodies will certainly evolve very differently in different individuals, and an additional difficult aspect will be to define rules for the administration of vaccine boosts, and whether those should be defined as a general rule (e.g. so many months after the initial vaccination), or individually, based on the monitoring the levels of neutralizing antibodies. For such an individually-based approach, HAT would seem to be a particularly appropriate solution since it is a very simple and cheap test based on the binding of antibodies to the RBD domain (Townsend et al. 2021), which are those with neutralizing activity. Furthermore, because the only reagent in HAT simply comprises small amounts of soluble IH4-RBD protein, the hemagglutination test can be very easily adapted to detect antibodies binding to variant forms of the virus (Jayathilaka et al. 2021).

Whilst HAT is not as sensitive as an anti-RBD ELISA or the Jurkat-S&R-flow test, this relatively low sensitivity may not really be a problem for using HAT to help decide when to revaccinate people since low levels of antibodies are unlikely to be fully protective. By performing titrations, the HAT assay can also provide a quantitative assessment of the levels of circulating antibodies, which have been shown to correlate strongly with sero-neutralisation titers (Lamikanra et al. 2021). In the current version of HAT, however, such a quantification can only be performed in a laboratory environment because it requires separation of the plasma or serum from the red blood cells. Making use of the Jurkat-S&R-flow test as a reference, we are currently in the process of completing work on a modified version of HAT that will be compatible with being performed pretty much anywhere, with no specialized equipment, and will provide a quantitative evaluation of the levels of antibodies in a single step (Joly et al. man in prep.).

**Detailed Contributions**

**Authors**

| Name | First name | ORCID | contributions |
|---|---|---|---|
| Maurel Ribes | Agnes | 0000-0002-7560-9502 | Collected and anonymized blood samples; scored HAT tests, corrected the manuscript |
| Bessière | Pierre | 0000-0001-5657-0027 | Provided hamster, cat and dog sera; performed SN; helped perform FACS assays, corrected the manuscript |
| Guéry | Jean Charles | 0000-0003-4499-3270 | Provided immunized mouse sera; corrected the manuscript |
| Joly Featherstone | Eloise | 0000-0001-9077-359X | Scored HAT tests |
| Bruel | Timothée | 0000-0002-3952-4261 | Provided Jurkat-R and -S cells; corrected the manuscript |
| Robinot | Rémy | 0000-0002-3651-0171 | Transduced, selected and grew Jurkat cells |
| Schwartz | Olivier | 0000-0002-0729-1475 | Provided Jurkat-R and -S cells |
| Volmer | Romain | 0000-0003-1591-3251 | Provided hamster sera |
| Abravanel | Florence | 0000-0002-1753-1065 | Provided cohort of sera and ELISA kits |
| Izopet | Jacques | 0000-0002-8462-3234 | Provided cohort of sera and ELISA kits; made suggestions for the manuscript |
| Joly | Etienne | 0000-0002-7264-2681 | Designed and funded the study; Performed the experiments; Wrote the paper. |

**Other contributors**

| | | |
|---|---|---|
| Townsend | Alain | Designed the IH4-RBD reagent and funded its production |
| Tiong | Tan | Produced the IH4-RBD reagent |
| Rijal | Pramilla | Produced the IH4-RBD reagent |
| Buchrieser | Julian | Generated lentiviral vectors |
| Porrot | Françoise | Grew Jurkat cells |
| Featherstone | Carol | Copy edited parts of the manuscript |

## Acknowledgements

The authors are extremely grateful to Alain Townsend, Tiong Kit Tan, Pramila Rijal, Julian Buchrieser and Françoise Porrot, who contributed the various materials detailed in the above table, and to Carol Featherstone for copy editing the manuscript. They also gratefully acknowledge the contributions of Marianne Navarra, for her help in setting up the agreement with the hospital; Miriam Pinilla, Karin Santoni and Etienne Meunier for the gift of Topro-3 and for performing the Mycoplasma tests; Emmanuelle Naser and Pénélope Viana from the IPBS flow cytometry facility for their assistance, and the very helpful staff of the Toulouse EFS.

Funding: The first part of this project was funded by a private donation. The second part was funded by the ANR grant HAT-field to EJ.

**Materials and Methods**

**Reagents**

Polyclonal anti-human and anti-mouse Igs secondary antibodies, all conjugated to Alexa-488, were from Jackson laboratories, and purchased from Ozyme (France) . Refs: anti-human Ig-GAM: 109-545-064, -G: 109-545-003, -A: 109-54-011, -M: 109-545-129; anti-mouse Ig-G: 115-545-003

Anti-cat IgG (F4262) and anti-dog IgG (F7884) secondary antibodies, both conjugated to FITC, were obtained from Sigma.

Anti-hamster IgG conjugated to Alexa 647 was obtained from Invitrogen (A-21451)

Anti RBD monoclonal antibodies: FI3A (site 1) and FD-11A (site 3) (Huang et al. 2021); C121 (site 2) (Robbiani et al. 2020); CR3022 (site 4) (ter Meulen et al. 2006); EY6A (site 4) (Zhou et al. 2020). All those were obtained using antibody-expression plasmids, as previously described (Townsend et al. 2021).

The 20/130 WHO reference serum was obtained from the NIBSC (Potters Bar, UK)

BSA Fraction V was obtained from Sigma (ref A8022).

PBS and tissue culture media were obtained from Gibco.

**Generation of Jurkat-S, Jurkat-R and Jurkat-G cell lines**
All three cell lines were obtained by means of lentiviral transduction.
pLV-EF1a-SARS-CoV-2-S-IRES-Puro was created by cloning a codon-optimized version of the SARS-CoV-2 S gene (GenBank: QHD43416.1) into the pLV-EF1a-IRES-Puro backbone (Addgene plasmid # 85132 ; http://n2t.net/addgene:85132 ; RRID:Addgene_85132) using BamHI and EcoRI sites.
The lentiviral vector for the expression of mCherry was obtained by replacing the GFP sequence of GFP by that of mCherry in the pCDH-EF1α-MCS*-T2A-GFP plasmid (https://systembio.com/shop/pcdh-ef1α-mcs-t2a-gfp-cdna-single-promoter-cloning-and-expression-lentivector/)
Lentiviral infectious supernatants were obtained after transient transfection of HEK cells with pLV- or pCDH-derived vectors, together with the packaging R8-2, and VSV-G plasmids. Supernatant was harvested 24h and 48h post transfection, passed through a 45 μm filter and stored in aliquots at -80°C. For the transduction of Jurkat cells, those were distributed in a 6 well plate at $1.5.10^6$ cell per well, in a volume of 500 μL of tissue cuture medium (RPMI, 10 % FCS, 1% PS, 2% Hepes), and 20 μL of the infectious supernatants were added, as well as 5 μL of Lentiblast premium (OZBiosciences). The plate was then spun at 1000g for 60 min at 32°C, before adding 2.5 ml of tissue culture medium and returning the plate to the 37°C incubator. Selection with puromycin was then performed at a concentration of 10μg/mL.
After a few days, the population of Jurkat cells thus obtained was stained for flow cytometry analysis using a highly reactive serum from a covid-19 patient, and it was found that most cells expressed the S protein, but at low to intermediate levels. To obtain a population that would express higher levels, we submitted this population to three successive rounds of sterile cell sorting using a FACSAria Fusion cell sorter (Beckton Dickinson), selecting each time the 5 % of cells with the brightest staining. Cells were placed back in culture and reamplified after each round of selection. At the time of the second round of sorting, the cell sorter was also used for single cell cloning, but all of the dozen clones obtained by this means showed lower expression that the sorted population. The expression of the S protein in the population of Jurkat cells obtained after these three rounds of sorting was found to remain expressed in all cells, and at similar levels, for more than 50 successive passages, over many weeks of continuous cell culture.

The Jurkat-R cells were obtained by successive transduction with the pCDH-GFP lentiviral vector described above, then with the empty pLV lentiviral vector, followed by selection with Puromycin at 10µg/mL. FACS analysis revealed that the mCherry fluorescent protein was expressed in 100 % of the cells of the population thus obtained, which therefore did not need to undergo any cell sorting.

After the initial selection process, both the Jurkat-S and Jurkat-R cell line were maintained in RPMI, 10 % FCS, 2 mM Glutamine, 1% PS and Puromycin at 2.5 µg/mL. The Jurkat-S and Jurkat-R cell lines were both checked for the absence of mycoplasma contamination using the HEK blue hTLR2 kit (Invivogen, Toulouse, France).

The Jurkat-G cell line was obtained by transduction with the Trip-GFP lentiviral vector (Zennou et al. 2000). Cell sorting was used to bring the 90% of the Jurkat cells that were expressing GFP after the initial transduction to 100 %. The cells were then kept in culture for several dozen passages in standard tissue culture medium with no detectable loss of GFP expression.

### FACS staining
Before experiments, cells in the cultures of both Jurkat-S and Jurkat-R cell lines were counted, and sufficient numbers harvested to have a bit more than $10^5$ cells of each per sample to be tested. Cells from both cell lines were then spun, and resuspended in their own tissue culture medium at a concentration of $2.2 \cdot 10^6$ cells/ mL before pooling equal volumes of the two.

Plasmas or sera to be tested were diluted 1/10, either in PBS or in PFN (PBS / 2% FCS / 200 mg/L sodium azide ). 10 µL of these 1/10 dilutions were then placed in U-bottom 96 well plates, before adding 90 ul per well of the Jurkat-S&R mix.

The plates were then incubated for 30 minutes at room temperature before placing them on ice for a further 30 minutes. As can be seen on the supplementary Figure 2, we have found that this initial incubation at room temperature results in a marked increase of the staining signals for most antibodies.

All subsequent steps were carried out in the cold, with plates and washing buffers kept on ice. After the primary staining, samples were then washed in PFN, with resuspending the cells by tapping the plate after each centrifugation, and before adding the next wash. After 3 washes, one drop (i.e. ca. 30 ul) of secondary fluorescent antibody diluted 1/200 in PFN was added to each well, and the cells resuspended by gentle shaking of the plates.

After an incubation of 60 min on ice, samples were washed two more times with cold PFN before transferring the samples to acquisition tubes in a final volume of 300 ul PFN containing 30 nM TO-PRO™-3 Iodide (Thermo Fischer Scientific, ref T3605).

This protocol was used for staining with just one pan-specific secondary antibody, i.e. for most of the samples of this study. For specific isotyping, i.e. for staining separately with either anti-IgG, -IgA or -IgM, as well as the pan-specific anti-Ig-GAM secondary antibody, we used double the quantities of cells and of serum or plasma for the primary step, and split the samples into 4 wells after the second wash, before proceeding to the subsequent steps as for the standard protocol.

The vast majority of the experiments for this study were analyzed on a FACScalibur flow cytometer controlled by the Cellquest pro software (Version 5.2, Beckton Dickinson), using the FL1 channel for Alexa-488 or FITC, the FL3 channel for m-Cherry, and the FL4 channel (with the 633 nm laser) for live gating with the TO-PRO™-3 live stain.

For the samples shown on supplementary Figure 1, double quantities were used (as for isotyping) so that the same samples could also be acquired on a Fortessa flow cytometer, controlled by the Diva software (Beckton Dickinson). After the samples had been run on both machines, an equivalent volume of PBS with 2% formaldehyde what added to what was left in the tubes, and those were stored at 4°C for 4 days before they were once again analyzed on both machines.

Post-acquisition analysis of all the samples was performed using the Flowjo software (version 10.7.1)

For the Jurkat-S&G-flow test, Jurkat-R cells were replaced by Jurkat-G cells, we used a secondary antibody conjugated to alexa 647, and TO-PRO™-3 Iodide was replaced by propidium iodide at 2 ug/ml final concentration. The samples were then analyzed on the same FACScalibur flow cytometer, using the FL1 channel for the GFP fluorescence, FL2 for live gating with propidium iodide, and FL4 for the Alexa-647 signals.

**Cost of the Jurkat-S&R-flow test**

The cost per sample of the Jurkat-S&R-flow test lies in large part with the price of the secondary antibodies used. Typically, one vial of secondary antibody costs about 150 € for 1ml, and using 30-50 ul at a 1/200 dilution will provide for at least 5000 samples. The cost of a polyclonal antibody is thus of the order or 3-5 cts per sample.

Each sample requires $2.10^5$ jurkat cells, which is roughly the amount obtained with 0.5 ml of standard tissue culture medium, which costs around 50 €/L, i.e. 2.5 cts/sample. All in all, if one adds the cost of TC, buffers for the washes and disposable plastics, we estimate that the cost per sample will be of the order of 10 cts.

The cost of access to a flow cytometer will be extremely variable between laboratories and institutes. If the cost of access is 20 € per hour, and one runs 200 samples per hour, this will add another 10 cts to the cost per sample.

**HAT tests**

HAT tests were performed using the IH4-RBD reagent diluted at 1 ug/ml in PBN rather than in straight PBS as originally described (Townsend et al. 2021), which results in a slight improvement of the HAT performances, and much improved stability of the IH4-RBD stocks ( Joly et al., man in prep.). PBN simply consists of PBS complemented with 1% BSA and 200 mg/L sodium azide. The main role of BSA is to prevent the IH4-RBD reagent from adsorbing onto the plastic of tubes and assay plates, and the azide prevents bacterial and fungal contaminations of the stocks. HAT tests were all performed in V-bottom 96 wells plates (Sarstedt, ref 82 1583).

For performing HAT on sera, two tubes of suspension of RBC from an O- donor were prepared at approximately 6µL of packed RBC per ml of PBN. The first tube was used to fill the negative control wells with 90 µL per well. The IH4-RBD reagent was added at 1.1 ug/ml to the second tube of RBC suspension and this was dispensed at 90 µL per test well. Sera to be tested were diluted 1/10 in PBS, and 10 µL of the dilutions were added to a control and a test well.

For performing HAT on whole blood samples, which were all leftover clinical samples collected in EDTA tubes (purple tops), 10 µL were diluted with 60 µL of PBS + 5 mM EDTA, and 10 µL of this 1/7 dilution were added to each of two wells containing either 90 µL of PBN for the negative controls, or 90 µL of PBN containing 1.1 µg/mL of IH4-RBD reagent for the test wells.

For both sera and whole blood samples, after 60 minutes incubation at room temperature, the V bottom plates were placed on a homemade light box tilted at an angle of approximately 10° from the vertical, and pictures taken with a mobile phone after approximately 20 seconds.

In an attempt to increase the sensitivity of the method by detecting partial hemagglutinations, the plates were then returned to a horizontal position for a further two hours, and the procedure of tilting the plate and taking pictures was repeated.

As a final step, to ensure that the RBCs in all the 267 blood samples of the second cohort were expressing glycophorin, 30 ul of the monoclonal antibody CR3022 diluted at 200 ng/ml in PBN were added to all samples and resuspended by pipetting with a multichannel pipet. The plates were tilted one more after another 60 minutes, and intense hemagglutination was seen in all 267 samples.

After transferring all the pictures to computer files, the hemagglutination tests were scored by three independent assessors, of which two were blinded.

The scoring system was as follows: A score of 1 was given only to those samples which showed complete hemagglutination after one hour. A score of 0.5 was given to samples which showed either partial hemagglutination after one hour, or partial or complete hemagglutination after the second incubation of 2 hours.

This revealed that, whilst there was 100% agreement between the three assessors for the scoring of complete hemagglutination, the scoring of partial hemagglutination proved to be much more subjective and variable, and only those samples which had been scored as partials by all three assessors were finally considered as bona fide partial hemagglutinations.

**ELISA**

The ELISA tests were performed using the WANTAI SARS-CoV-2 Ab ELISA (ref WS-1096), according to the manufacturer's instructions. This commercial kit allows detection of all human antibody isotypes recognizing a recombinant RBD domain of the SARS-Cov-2 virus.

The ELISA tests for the cohort of 121 sera were performed in the virology department of the Toulouse hospital, as described previously (Abravanel et al. 2020).

The series of ELISA tests for the cohort of 267 plasmas were performed at the IPBS, with the washes being performed by hand rather than by an automated machine, and the 450/625 ODs read by a µQuant plate reader. All the positive samples of this cohort, as well as a set of randomly selected negatives, were submitted to a repeat of the assay, which showed excellent reproducibility.

**Human samples**

Purpan cohort: The collection of 2019 (negative) and SARS-CoV-2 infected (positive) patients was obtained from the laboratory of virology of the Purpan hospital, as previously described (Abravanel et al. 2020).

Rangueil Cohort: The samples were routine care residues from random patients, anonymized within 48 hours of collection, regardless of gender or hospitalization reason, collected in the course of a month between the end of January and the end of February 2021. The Covid status (PCR or positive serology) was unknown to the person performing the Jurkat-S&R-flow, ELISA and HAT experiments. Those whole blood samples were kept at room temperature until being used for HAT assays within 24 hours of obtaining them from the hospital. In trial experiments, we had found that such samples could be stored for up to 5 days without any noticeable difference in the performance of the HAT test.

After the whole blood samples had been used for HAT assays, the tubes were then spun, and the plasmas harvested into fresh tubes (to which sodium azide was added at a final concentration of 200 ug/ml). Those harvested plasmas were kept at 4°C until they were used to perform the Jurkat-S&R-flow tests and ELISA tests.

Capillary blood: one of the authors of this study collected 50 µL of his own blood by finger pricking, using disposable lancets (Sarstedt ref 85.1016 ) on various days during the course of his vaccination regimen with the Pfizer/BioNTech vaccine. At the time of collection, the blood samples were diluted with 200 µL of PBS + 5mM EDTA (Since whole blood is roughly 50% RBCs and 50% plasma, this 1/5 dilution of the blood actually corresponds to a 1/10 dilution of the plasma). The plasma was then separated from the RBCs after centrifugation, placed in another tube with sodium azide, and stored at 4°C until the day of the assay.

**Experiments on mouse sera**

Virus preparation and inactivation

SARS-CoV2 was grown on Vero E6 cells (ATCC) in DMEM (Dutscher) supplemented with 100 U/mL penicillin, 100 µg/ml streptomycin (Invitrogen), and 2% heat inactivated fetal bovine serum (Sigma). Viral stocks were propagated in 300 cm$^2$ flasks (Dutscher) in which 10$^3$ tissue culture infectious dose 50 (TCID$_{50}$) were inoculated in 100 ml of medium for 3 days at 37°C with 5% CO$_2$. Culture supernatants containing the viral stocks were harvested and inactivated with BPL (Fischer) overnight at 4°C. Virus was then concentrated by ultracentrifugation at 25 000 rpm (for 2 hours at 4°C) on a 20% sucrose cushion in Ultra-Clear centrifuge tubes (SW-32 Ti rotor, Beckman). Pellets were resuspended in PBS, protein concentration was quantified by BCA Protein Assay kit (Pierce). BPL-inactivated SARS-CoV2 stocks at protein concentrations ranging from 1 to 5 µg/µL were stored at -80°C until use.

Mouse immunization

Mice were injected intra-peritoneally with 15 µg of BPL-SARS-CoV2 in 250 µl of PBS, and then challenged with the same amount at day 62. Mice were euthanized and serum was collected at day 14 post-primary immunization and at day 7 post-secondary challenge. Those experiments were conducted within the scope of the APAFIS licence n° 15236 delivered to Jean Charles Guéry.

**Experiments on Cat and Dog sera**

Serum samples were collected from cats and dogs belonging to owners who developed COVID-19-like symptoms and subsequently tested positive for SARS-CoV-2 infection by RT-qPCR. , at least one month after the owners' recoveries. Samples and data collections were conducted according to the guidelines of the Declaration of Helsinki, and approved by the Ethics Committee Sciences et Santé Animale n◦115 (protocol code COVIFEL approved on 1 September 2020, registered under SSA_2020_010).

<u>Sero-neutralisation assay</u>

Serum samples and controls were heat-inactivated at 56 °C for 30 min, serially diluted in DMEM starting at 1:10, mixed with an equal volume of SARS-CoV-2 stock (previously amplified and titrated on Vero-E6 cells and diluted in DMEM to contain 2000 $TCID_{50}$/ml), incubated for 2 hours at 37 °C, and 100 μL transferred to tissue-culture 96 well plates plated with 12.000 Vero-E6 cells per well the day before the assay, in DMEM complemented with 10% of heat-inactivated fetal bovine serum and 1% of penicillin-streptomycin at 37 °C with 5% of CO2 (medium was removed before adding the virus-serum dilutions). After 60 minutes at 37°C, the virus-serum dilutions were removed before adding DMEM complemented with 2% of heat-inactivated fetal bovine serum and 1% of penicillin-streptomycin. Cells were then incubated for 72 h at 37 °C with 5% of CO2. Individual wells were then screened by eye under the microscope for cytopathic effects. Monoclonal anti-SARS-CoV CR3022 antibody (BEI Resources, NIAID, NIH) was used as a positive control. PBS was used as a negative control. Experiments were carried out in a biosafety level 3 facility at the National Veterinary School of Toulouse.

**Experiments on virally-infected hamsters**

Eight week-old female Syrian golden hamsters (*Mesocricetus auratus*, strain RjHan:AURA) from Janviers's breeding Center (Le Genest, St Isle, France) were housed in an animal-biosafety level 3 (A-BSL3), with *ad libidum* access to water and food. Animals were anesthetized with isoflurane and inoculated intranasally with $10^4$ $TCID_{50}$ units of UCN1 SARS-CoV-2 strain split in 20μL in each nostril. SARS-CoV-2 strain UCN1 was amplified as described previously and used at passage 2 (Monchatre-Leroy et al. 2021; Bessière, Wasniewski, et al. 2021). Non-infected animals received the equivalent amount of PBS. Animals were weighted and clinically monitored daily. Six animals from each group were anesthetized and euthanized by exsanguination at 15 dpi. Sera were harvested and aliquots were kept frozen until they were needed for the experiments described here.

The animal experimentation protocols complied with the regulation 2010/63/CE of the European Parliament and of the council of 22 September 2010 on the protection of animals used for scientific purposes. These experiments were approved by the Anses/ENVA/UPEC ethic committee and the French Ministry of Research (Apafis n°24818-2020032710416319).

**Ethical statement**

All sera from the first cohort, and whole blood samples from the second cohort, were obtained from the Toulouse hospital, where all patients give, by default, their consent for any biological material left over to be used for research purposes after all the clinical tests requested by doctors have been duly completed. Material transfer was done under a signed agreement (CNRS n° 227232, CHU n° 20 427 C). This study was declared and approved by the governing body of the Toulouse University Hospital with the agreement number RnIPH 2021-99, confirming that ethical requirements were fully respected.

Using such materials, we did not really have control over how representative our cohorts may be, but this allowed us to circumvent the very stringent rules set by French laws regarding the use of human materials for research (RIPH, Loi Jardé). Under those rules, setting up a clinical trial involving biological materials of human origin would require many months of administrative procedures and paperwork, as well as several hundreds of thousands of euros, which we did not have access to.

For the same reason, the blood samples for the experiment shown on Figure 3A were collected by one of the authors by simple finger-pricking. He did so on his own free will, without the intervention of anybody else, thus circumventing the need for ethical approval.

# Supplementary material

## Figure S1

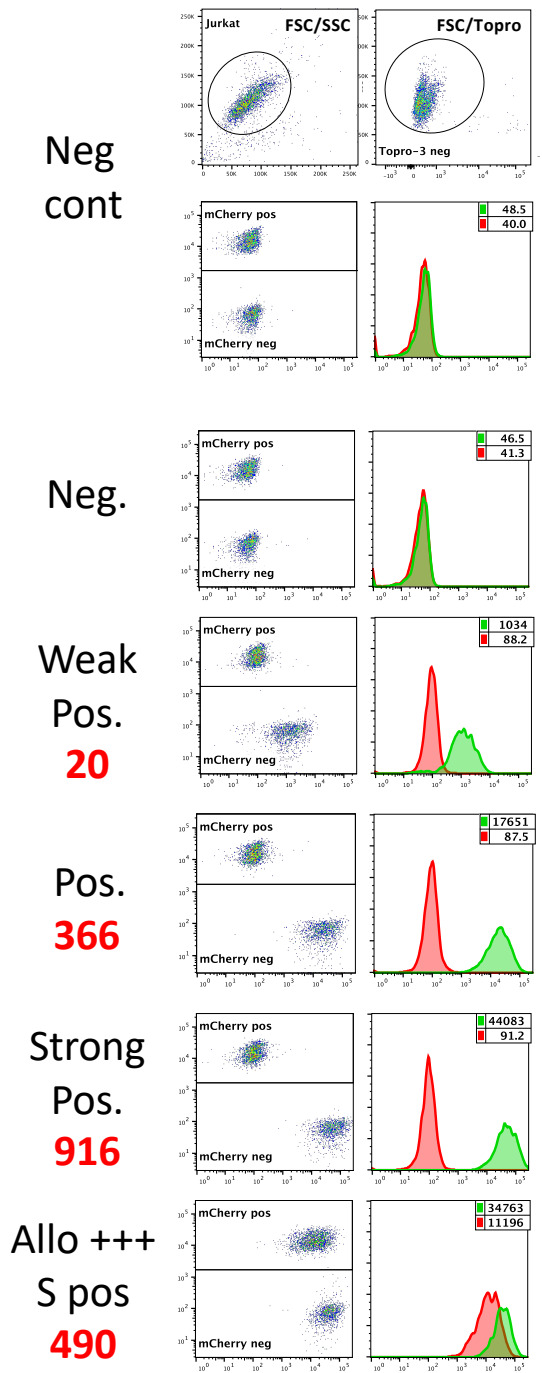
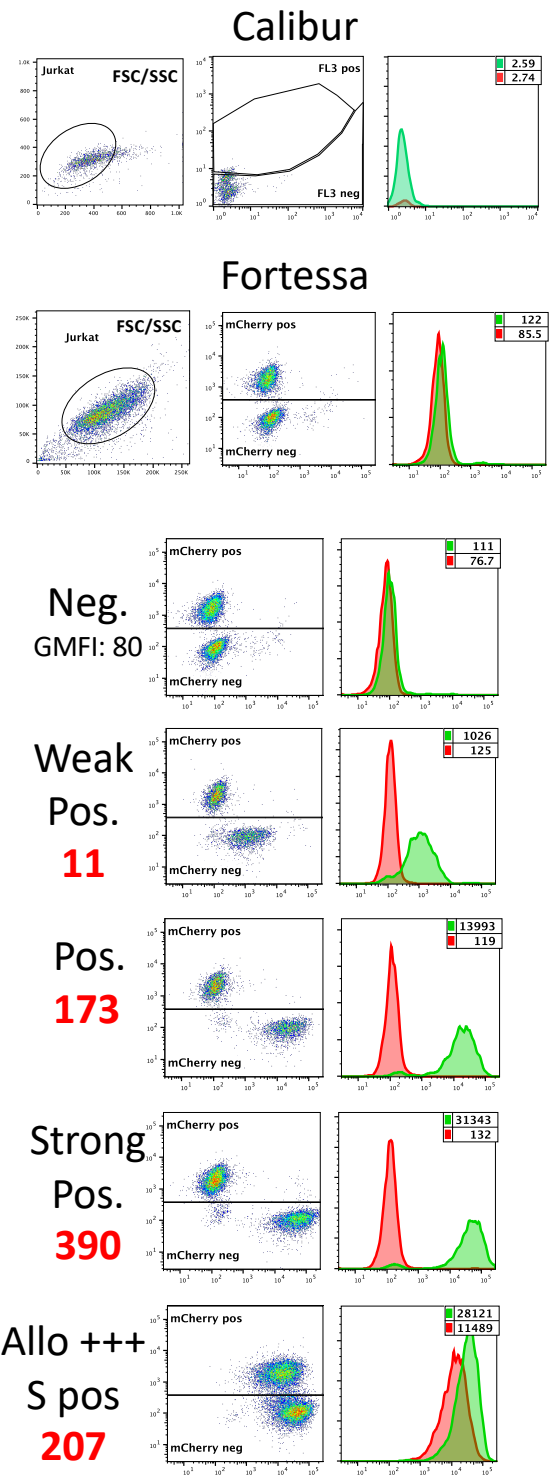

Fixed 1% formaldehyde, 4 days @ 4°C

Calibur / Fortessa

Neg cont / Neg. (GMFI: 80 / 111 / 76.7)
Weak Pos. (20 / 11)
Pos. (366 / 173)
Strong Pos. (916 / 390)
Allo +++ S pos (490 / 07)

**Figure S1**: <u>Using a flow cytometer with a 561 nm Yellow-Green laser improves the mCherry signals</u>

The same samples shown on Figure 1 were also analyzed on a Fortessa flow cytometer. The 561 nm yellow-green laser was used for the excitation of the mCherry fluorescent protein, which led to much brighter signals, and thus to much easier separation of the Jurkat-R (mCherry pos) and Jurkat-S (mCherry neg) populations than when the samples were acquired on a FACScalibur.

As in figure 1, the numbers in the upper right corners of the histogram overlays plots correspond to the GMFI of the two histograms (Red: Jurkat-R; green: Jurkat-S), and the big red numbers to the left of the plots to the RSS. Of note, although the GMFI values were much higher on the Fortessa than those recorded on the FACScalibur, the RSS values of the various samples were all very similar between the two machines.

After the samples had been run on both flow cytometers, those were fixed in 1% formaldehyde by adding an equal volume of PBS + 2% formaldehyde to the remainder of each sample. The tubes were then stored à 4°C for 4 days before re-analyzing them on both cytometers. As can be seen on the right hand side of the figure, the red fluorescence of the mCherry was still detectable on the Fortessa after this procedure of fixation, despite the intensity of the signals having dropped by about 7 fold. On the FACScalibur, however, this drop of the mCherry fluorescent signals precluded the separation of the Jurkat-S from the Jurkat-R cells. If samples are to be fixed, performing the Jurkat-R&S-flow test will thus require access to a flow cytometer equipped with a 561 nm Yellow-Green laser. As an alternative, we are currently exploring the possibility of transforming the Jurkat-S&R-flow test into a Jurkat-S&G-flow test, where the negative control cells would be expressing GFP, which is excited at 488 nm, and using secondary antibodies conjugated to red fluorochromes such as Alexa647, excited by a 633 nm red laser, for the detection of the primary antibodies.

Of note, in the dot plots of the Positive and Strong Positive fixed samples, faint clouds of FL1 negative cells can be seen in the mCherry neg window. Those presumably correspond to Jurkat-R cells which were dead before the fixation step, and from which the mCherry had thus leaked out. When analyzing populations of live cells, such cells are gated out with the To-pro 3 live stain, but this no longer works for population of fixed cells. This small problem could probably be circumvented using a fixable dead cell staining dye such as the LIVE/DEAD Fixable Far Red stain (ThermoFischer L34973).

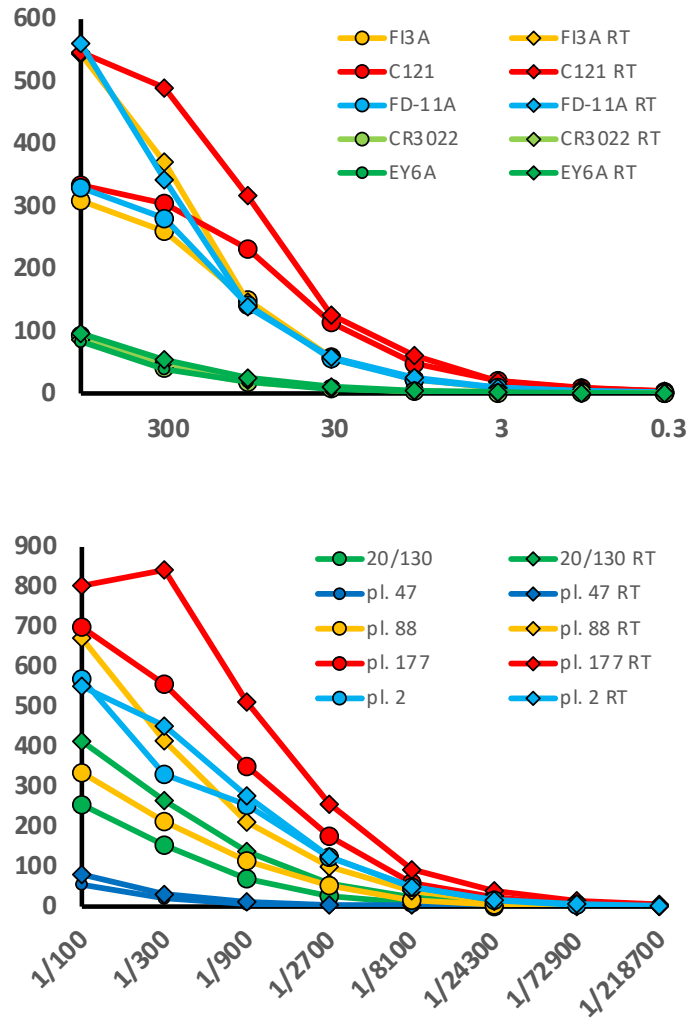

**Figure S2**: <u>Incubation of the samples at room temperature for the first part of the primary staining step markedly improves the staining signals for most antibodies.</u>

For a large fraction of monoclonal and polyclonal antibodies reacting with the Spike protein expressed at the surface of cells, we have repeatedly observed that a step of incubation at room temperature can result in a marked improvement of the staining signals compared to performing the staining continuously on ice.

For this experiment, we selected a panel of 5 monoclonal antibodies reacting with different sites of the spike RBD domain, a panel of 4 plasmas reacting with various intensities, as well the 20/130 reference serum.

For all the samples, we performed eight steps of 3-fold dilutions, and placed 10 ul of the diluted samples in two parallel U bottom 96 well plates. The first plate was kept at room temperature (RT) whilst the other one was placed on ice. The Jurkat-S&R mix was then added to the wells of the plate at room temperature. The tube was then placed on ice for a few minutes, before distributing the mix in the wells of the second plate. After 30 minutes, the plate which had been sitting at room temperature was also placed on ice, and incubated for another 30 minutes. The rest of the staining procedure was then exactly the same for the two plates, all performed in the cold as described in M&M.

|  | Sample ID | IgGAM | | | IgG | | | IgA | | | IgM | | | G+A+M | Fraction IgG signal | Fraction IgA signal | Fraction IgM signal |
|---|---|---|---|---|---|---|---|---|---|---|---|---|---|---|---|---|---|
|  |  | GMFI Jurkat-R | GMFI Jurkat-S | specific stain (J-S - J-R) | GMFI Jurkat-R | GMFI Jurkat-S | specific stain (J-S - J-R) | GMFI Jurkat-R | GMFI Jurkat-S | specific stain (J-S - J-R) | GMFI Jurkat-R | GMFI Jurkat-S | specific stain (J-S - J-R) |  |  |  |  |
|  | neg | 1.87 | 2.01 | 0.14 | 1.95 | 1.99 | 0.04 | 1.79 | 2.13 | 0.34 | 1.89 | 1.98 | 0.09 | 0.47 |  |  |  |
| Sera | 30/120 | 3.19 | 760 | 756.81 | 3.59 | 763 | 759.41 | 2.11 | 63.2 | 61.09 | 2.74 | 640 | 637.26 | 1457.76 | 0.52 | 0.04 | 0.44 |
|  | <J14-3 | 5.1 | 51.6 | 46.5 | 7.47 | 57.2 | 49.73 | 3.08 | 29.1 | 26.02 | 2.55 | 8.65 | 6.1 | 81.85 | 0.61 | 0.32 | 0.07 |
|  | <J14-29 | 4.59 | 738 | 733.41 | 5.16 | 919 | 913.84 | 2.77 | 122 | 119.23 | 2.54 | 203 | 200.46 | 1233.53 | 0.74 | 0.10 | 0.16 |
|  | <J14-32 | 4.71 | 1692 | 1687.29 | 5.25 | 2350 | 2344.75 | 4.01 | 610 | 605.99 | 2.56 | 402 | 399.44 | 3350.18 | 0.70 | 0.18 | 0.12 |
|  | <J14-35 | 4.65 | 198 | 193.35 | 4.82 | 209 | 204.18 | 1.97 | 52.6 | 50.63 | 3.14 | 13.1 | 9.96 | 264.77 | 0.77 | 0.19 | 0.04 |
|  | <J14-109 | 6.52 | 852 | 845.48 | 6.23 | 1082 | 1075.77 | 2.61 | 179 | 176.39 | 4.55 | 126 | 121.45 | 1373.61 | 0.78 | 0.13 | 0.09 |
|  | <J14-130 | 8.7 | 150 | 141.3 | 9.46 | 151 | 141.54 | 3.28 | 45.7 | 42.42 | 5.25 | 58.1 | 52.85 | 236.81 | 0.60 | 0.18 | 0.22 |
|  | >J14-40 | 4.84 | 54.9 | 50.06 | 6.07 | 54.3 | 48.23 | 2.18 | 12 | 9.82 | 2.76 | 29.1 | 26.34 | 84.39 | 0.57 | 0.12 | 0.31 |
|  | neg-2213 | 10.5 | 33.3 | 22.8 | 12.3 | 43.3 | 31 | 3.21 | 6.58 | 3.37 | 6.12 | 8.77 | 2.65 | 37.02 | 0.84 | 0.09 | 0.07 |
|  | neg-2649 | 19.2 | 41.4 | 22.2 | 24.7 | 56.2 | 31.5 | 6.32 | 9.59 | 3.27 | 6.15 | 9.62 | 3.47 | 38.24 | 0.82 | 0.09 | 0.09 |
| Plasmas | 2 | 438 | 1427 | 989 | 577 | 1846 | 1269 | 59.4 | 469 | 409.6 | 739 | 1333 | 594 | 2272.60 | 0.56 | 0.18 | 0.26 |
|  | 18 | 7.82 | 84.9 | 77.08 | 11.3 | 101 | 89.7 | 2.67 | 33.4 | 30.73 | 4.89 | 12.7 | 7.81 | 128.24 | 0.70 | 0.24 | 0.06 |
|  | 19 | 2.02 | 2.2 | 0.18 | 2.07 | 2.15 | 0.08 | 1.82 | 2.18 | 0.36 | 1.83 | 1.96 | 0.13 | 0.57 |  |  |  |
|  | 36 | 4.53 | 88.7 | 84.17 | 5.01 | 90.2 | 85.19 | 2.38 | 48.5 | 46.12 | 3.61 | 10.4 | 6.79 | 138.10 | 0.62 | 0.33 | 0.05 |
|  | 47 | 3.47 | 139 | 135.53 | 3.65 | 141 | 137.35 | 2.15 | 47.1 | 44.95 | 2.8 | 42.2 | 39.4 | 221.70 | 0.62 | 0.20 | 0.18 |
|  | 56 | 9.03 | 25.7 | 16.67 | 9.48 | 26.8 | 17.32 | 2.56 | 17.2 | 14.64 | 7.16 | 11.2 | 4.04 | 36.00 | 0.48 | 0.41 | 0.11 |
|  | 57 | 5.84 | 25.5 | 19.66 | 7.04 | 26.7 | 19.66 | 2.17 | 8.87 | 6.7 | 4.55 | 13.9 | 9.35 | 35.71 | 0.55 | 0.19 | 0.26 |
|  | 73 | 3.87 | 130 | 126.13 | 3.61 | 159 | 155.39 | 1.91 | 4.35 | 2.44 | 3.1 | 4.64 | 1.54 | 159.37 | 0.98 | 0.02 | 0.01 |
|  | 80 | 2.5 | 82 | 79.5 | 3.07 | 99.1 | 96.03 | 2.06 | 2.75 | 0.69 | 2.21 | 2.56 | 0.35 | 97.07 | 0.99 | 0.01 | 0.00 |
|  | 83 | 3.13 | 1674 | 1670.87 | 3.69 | 1900 | 1896.31 | 2.21 | 542 | 539.79 | 2.4 | 629 | 626.6 | 3062.70 | 0.62 | 0.18 | 0.20 |
|  | 88 | 3.74 | 1110 | 1106.26 | 4.18 | 1297 | 1292.82 | 2.51 | 159 | 156.49 | 3.03 | 632 | 628.97 | 2078.28 | 0.62 | 0.08 | 0.30 |
|  | 108 | 3.53 | 7.65 | 4.12 | 4.22 | 9.66 | 5.44 | 2.08 | 2.61 | 0.53 | 2.6 | 3.58 | 0.98 | 6.95 |  |  |  |
|  | 123 | 54.5 | 75.6 | 21.1 | 51.6 | 71.4 | 19.8 | 4.22 | 6.14 | 1.92 | 43.5 | 61.1 | 17.6 | 39.32 | 0.62 | 0.08 | 0.30 |
|  | 161 | 2.91 | 5.84 | 2.93 | 6.17 | 11.1 | 4.93 | 2.08 | 2.59 | 0.51 | 2.09 | 2.25 | 0.16 | 5.60 |  |  |  |
|  | 166 | 101 | 162 | 61 | 109 | 176 | 67 | 27.2 | 35.3 | 8.1 | 46.5 | 68.9 | 22.4 | 97.50 | 0.69 | 0.08 | 0.23 |
|  | 174 | 3.11 | 190 | 186.89 | 4.43 | 228 | 223.57 | 2.45 | 50.1 | 47.65 | 2.19 | 42.3 | 40.11 | 311.33 | 0.72 | 0.15 | 0.13 |
|  | 177 | 7.9 | 1945 | 1937.1 | 8.04 | 2303 | 2294.96 | 2.32 | 581 | 578.68 | 7.58 | 699 | 691.42 | 3565.06 | 0.64 | 0.16 | 0.19 |
|  | 227 | 4.12 | 77 | 72.88 | 5.31 | 51.4 | 46.09 | 2.83 | 6.19 | 3.36 | 3 | 73.8 | 70.8 | 120.25 | 0.38 | 0.03 | 0.59 |
|  | 232 | 11 | 131 | 120 | 12.6 | 169 | 156.4 | 6.21 | 9.13 | 2.92 | 4.84 | 6.83 | 1.99 | 161.31 | 0.97 | 0.02 | 0.01 |
|  | 241 | 4.11 | 2411 | 2406.89 | 4.61 | 2860 | 2855.39 | 2.6 | 608 | 605.4 | 2.83 | 47.8 | 44.97 | 3505.76 | 0.81 | 0.17 | 0.01 |
|  | 247 | 13.8 | 23.3 | 9.5 | 21.6 | 31.7 | 10.1 | 8.87 | 15.7 | 6.83 | 3.5 | 5.88 | 2.38 | 19.31 | 0.52 | 0.35 | 0.12 |
|  | 253 | 3.98 | 28.2 | 24.22 | 5.22 | 33 | 27.78 | 2.43 | 10.7 | 8.27 | 2.43 | 2.9 | 0.47 | 36.52 | 0.76 | 0.23 | 0.01 |
|  | 260 | 23.2 | 33.9 | 10.7 | 30.8 | 44.1 | 13.3 | 29.4 | 43 | 13.6 | 5.98 | 7.61 | 1.63 | 28.53 | 0.47 | 0.48 | 0.06 |

**Table S1**: Example of isotyping performed on a representative set of various serum and plasma samples.

As can be seen on the right hand side portion of the table, comparison of samples from different patients shows that there can be large variations in the proportions of Ig-G,-A and -M. Because polyclonal secondary antibodies will not all react with their targeted primary antibodies with the same efficiency, however, the actuals signals for the various subclasses of antibodies should not be taken as a true reflection of the amounts of each subclass of antibody.

If a more quantitative evaluation was needed, monoclonal secondary antibodies could be used instead of polyclonal reagents, making sure that all those isotypes-specific secondary monoclonal antibodies were being used well above the saturating concentration, and conjugated to similar amounts of the same fluorochrome. This would, however, increase the cost of the procedure considerably.